\documentclass[twocolumn,showpacs,aps,prl,amsmath,amssymb]{revtex4}

\usepackage{graphicx}

\begin{document}

\title{Energy gap of the bimodal two-dimensional Ising spin glass}

\author{W. Atisattapong}
\affiliation{Department of Mathematics, Faculty of Science,
Mahidol University, Rama 6 Road, Bangkok 10400, Thailand}

\author{J. Poulter}
\affiliation{Department of Mathematics, Faculty of Science,
Mahidol University, Rama 6 Road, Bangkok 10400, Thailand}

\date{\today}

\begin{abstract}
An exact algorithm is used to compute the degeneracies of the excited states
of the bimodal Ising spin glass in two dimensions. It is found that the
specific heat at arbitrary low temperature is not a self-averaging quantity
and has a distribution that is neither normal or lognormal. Nevertheless, it
is possible to estimate the most likely value and this is found to scale as
$L^3 T^{-2} \exp(-4J/kT)$, for a $L \times L$ lattice. Our analysis also explains,
for the first time, why a correlation length $\xi \sim \exp(2J/kT)$ is
consistent with an energy gap of $2J$. Our method allows us to obtain results for up to $10^5$ disorder realizations with $L\leq64$. Distributions of second and third excitations are also shown.

\end{abstract}


\maketitle

In spite of its comparative simplicity, the two-dimensional bimodal
short-range Ising spin glass model remains an interesting source of controversy. Although it is now widely accepted that the spin glass only exists at zero temperature \cite {HY01,H01}, the nature of excitations, in particular the energy gap with the first excited state, has commanded much interest in the literature.

The bimodal model has bond (nearest-neighbor) interactions of fixed magnitude
$J$ and random sign. If we think of an infinite square lattice, without open
boundaries, it is easy to appreciate that any finite number of spin flips
cannot result in an excitation energy of less than $4J$. Nevertheless, some
20 years ago, Wang and Swendsen \cite {WS88} gave credible evidence that the
energy gap with the first excited state should be $2J$ in the thermodynamic
limit. These excitations must involve an infinite number of spin flips. The
issue here is the noncommutativity of the zero-temperature and thermodynamic
limits. In such a situation it is imperative to perform the thermodynamic
limit first.

Support for the $2J$ energy gap has included studies that have involved
exact computations of partition functions \cite {LGM04}, a worm algorithm
\cite {W05} and Monte Carlo simulation \cite {KLC05}. Challenges to $2J$
have also appeared. Saul and Kardar \cite {SK93} maintained that the energy
gap should be $4J$ as naive analysis suggests. More recently \cite {JLM06},
it has been reported that the specific heat should follow a power law in
temperature. In particular, it was proposed that the critical exponents
must be the same as those for the model with a Gaussian distribution of
bond interactions, indicating universality with respect to the type of
disorder. Fisch \cite {Fisch}, using a steepest descent approximation,
has also suggested power law behaviour albeit with a different exponent.

An important quantity, intimately involved with studies of the thermodynamic
limit at fixed temperature, is the correlation length. Essentially this
measures the spatial extent of the influence of one spin on others.
The correlation length is infinite at a critical temperature. For the
spin glass model of interest here, the correlation length has been
determined by reliable Monte Carlo techniques \cite {H01,KL04} to be
$\xi \sim \exp(2J/kT)$. This is in agreement with Ref. \cite {SK93} and
consistent with a qualitative study \cite {Ratee}. If this is true then
hyperscaling predicts that the energy gap should be $4J$. A $2J$ gap would be
consistent with $\xi \sim \exp(J/kT)$ as proposed in Ref. \cite {KLC05}.
Another scenario is power law behaviour \cite {JLM06}.

A good current review of the issues involved here has been given in
Ref. \cite {KLC07}. The power law behaviour of Ref. \cite {JLM06}
is discussed in the light of new Monte Carlo data. The conclusion is that
the suggested universality cannot be reliably proven with the computational
facilities currently available. Further, it is stated that extrapolation
of the data of Ref. \cite {KLC05} to zero temperatures may not be
plausible. The main message of this letter is to argue, for the first time,
why a correlation length $\xi \sim \exp(2J/kT)$ is perfectly consistent
with an energy gap $2J$, in apparent violation of hyperscaling. 

We have performed exact calculations of the degeneracies of excited states
at a fixed arbitrarily low temperature. Each disorder realization consisted
of a frustrated $L \times L$ patch with periodic boundary conditions in one
dimension, embedded in an infinite unfrustrated environment in the second
dimension. This choice of boundary conditions, with even $L$, definitely does
not allow any first excitation with energy gap less than $4J$. There are no
open boundaries and no diluted bonds. A $2J$ energy gap can only arise in the
thermodynamic limit.

Any planar Ising model is isomorphic to a system of interacting fermions. The
Pfaffian formalism \cite {GH64} is particularly convenient for disordered
systems \cite {B82,BP91}. Each bond is decorated with two fermions, one
either side, so that a square plaquette has four; left, right,
top and bottom. The partition function is given by $Z \sim (\det D)^{1/2}$
where $D$ is a real skew-symmetric $4N \times 4N$ matrix for a lattice with $N$
sites. Perturbation theory is used to determine ground state properties.
Basically we require the low-temperature behaviour of the defect (meaning
zero at zero temperature) eigenvalues of $D$. Exactly, $D=D_{0}+\delta D_{1}$
where $\delta=1-t$ with $t=\tanh(J/kT)$. The temperature dependence appears
in $\delta$ only. $D_{0}$ is singular in the ground state and has eigenvectors $\mid d \rangle$
corresponding to zero eigenvalue, that is $D_{0}\mid d\rangle =0$,
localized on each frustrated plaquette. 
These
eigenvectors form the subbasis for the perturbation theory. 

At first order, the matrix $D_{1}$, which is $2 \times 2$ block diagonal across bonds, is diagonalized in the subbasis. To continue at second
order requires the continuum Green's function \cite {BP91,PB05} $G_{c}=(1-P)g_{c}(1-P)$ where $P = \sum_d \mid d\rangle\langle d\mid$ and $g_{c} = g_{c1} + g_{c2}$. The first term $g_{c1}$ is $4 \times 4$ block
diagonal in the four fermions in each plaquette and allows us to connect
frustrated plaquettes across two bonds. At second order we diagonalize
$D_{2} = D_{1} g_{c1} D_{1}$. The matrix $g_{c2}$ is, just like $D_{1}$,
$2 \times 2$ block diagonal across bonds and only relevant for
excited states. For higher orders we require Green's functions $G_{r}$ constructed
at previous orders. At third order the matrix to be diagonalized is
$D_{3}=D_{2}(1+G_{1}D_{1})g_{c1}D_{1}$
and, generally at arbitrary order
$D_{r}=D_{r-1}(1+G_{r-2}D_{r-2})\cdots(1+G_{1}D_{1})g_{c1}D_{1}$
until all degeneracy is lifted. Here we define, for $r \geq 1$,
\begin{equation}
G_r = \sum_{i=1}^{N(r)} \frac{\mid\alpha_r^i\rangle\langle\beta_r^i\mid-\mid\beta_r^i\rangle\langle\alpha_r^i\mid}{\epsilon_r^i}
\label{e:GR}
\end{equation}
where $D_r\mid\alpha_r^i \rangle=-\epsilon_r^i\mid\beta_r^i\rangle$ and $D_r\mid\beta_r^i \rangle=\epsilon_r^i\mid\alpha_r^i\rangle$ with $\epsilon_r^i\neq0$ and there are $N(r)$ such pairs at order $r$. Note that $G_r$ is real here, as are all matrices; in contrast to Ref. \cite{BP91} where they are imaginary.

The internal energy of the system is 
\begin{equation}
U=-J\sum_{\langle ij \rangle}\zeta_{ij}\langle \sigma_i\sigma_j \rangle
\label{e:U}
\end{equation}
where $\sigma_i$ is an Ising spin, $\zeta_{ij}$ represents the sign of the bond and the nearest-neighbor correlation function can be written \cite {B82}
\begin{equation}
\zeta_{ij}\langle \sigma_i\sigma_j \rangle=t-(1-t^2)\mathcal{G}^{-+}
\label{e:Cor}
\end{equation}
where $\mathcal{G}^{-+}$ means the matrix element of $\mathcal{G}=-D^{-1}$ 
between the two nodes decorating the bond $\langle ij \rangle$ \cite{bond_basis}.

The full Green's function $\mathcal{G}$ can be expanded for a finite system in powers of $\delta$
\begin{equation}
\mathcal{G}=\sum_{m=-\infty}^{r_{max}} \delta^{-m}K_m
\label{e:Gk}
\end{equation}
where $r_{max}$ is the highest order of perturbation theory required. It is obvious that $K_m$, for $m>1$, has no direct physical meaning. We provide here a brief outline of how the perturbation theory for the Green's function allows us to expand the internal energy exactly.

Equating powers of $\delta$ in $\mathcal{G}D=-1$, we obtain (for $m<r_{max}$)
\begin{equation}
D_0K_m+D_1K_{m+1}=-\delta_{m0}.
\end{equation}
Since $PD_{0}=0$, we can show that for $m\leq0$,
\begin{equation}
K_m=(1+K_1D_1)g_c(\delta_{m0}+D_1K_{m+1}).
\end{equation}
It also can be proven that $(1+K_1D_1)$ is idempotent, that is  
\begin{equation}
(1+K_1D_1)^2=(1+K_1D_1).
\label{e:id}
\end{equation}

Since both $g_{c2}$ and $D_1$ are bond diagonal and $g_{c2}D_1=\frac{1}{2}$ \cite{PB05}, we get $K_0=(1+K_1D_1)g_{c1}(1+D_1K_1)+g_{c2}+\frac{1}{2}K_1$, and for $m<0$
\begin{equation}
K_m=((1+K_1D_1)g_{c1}D_1+\frac{1}{2})K_{m+1}.
\end{equation}

Substituting $t=1-\delta$ and the Green's function in Eq. (\ref{e:Cor}), the correlation function can be expressed in terms of $K_1$ as
\begin{align}
\zeta_{ij}\langle \sigma_i\sigma_j \rangle&=(1-2K_1^{-+})-2\sum_{m=1}^\infty\delta^m[((1+K_1D_1)g_{c1}D_1\nonumber\\
&+\frac{1}{2})^{m-1}(1+K_1D_1)g_{c1}(1+D_1K_1)]^{-+}
\label{e:UK}
\end{align}

We can now use the binomial theorem to expand in $\exp(-2J/kT)$ instead of $\delta$, which eliminates the explicit effect of $g_{c2}$, and the internal energy can be expressed as
\begin{equation}
U=\sum_{m=0}^{\infty}e^{-2Jm/kT}U_m
\end{equation}
where $U_0=-2JN+2J\sum_{\langle ij \rangle}K_1^{-+}$ and for $m>0$
\begin{align}
U_m=&2^{m+1}J\sum_{\langle ij \rangle}[\left((1+K_1D_1)g_{c1}D_1\right)^{m-1}\nonumber\\
&(1+K_1D_1)g_{c1}(1+D_1K_1)]^{-+}.
\label{e:Um}
\end{align}

Since $D_1^{+-}=-1$, it is obvious that $2\sum_{\langle ij \rangle}A^{-+}=-\sum_{\langle ij \rangle}(D_1A)^{++}-\sum_{\langle ij \rangle}(AD_1)^{--}=-\mathrm{Tr}(AD_1)$ for any skew-symmetric matrix $A$.
We can then show under the trace that for $m>0$, 
\begin{equation}
U_m=-2^{m}J\mathrm{Tr}[((1+K_1D_1)g_{c1}D_{1})^m(1+K_1D_1)].
\label{e:Umge}
\end{equation}
This can be rewritten using the idempotence relation (Eq. (\ref{e:id})) to get
\begin{equation}  
U_m=-2^{m}J\mathrm{Tr}(D_1g_{c1}(1+D_1K_1))^m.
\label{e:trk}
\end{equation}

Finally, with recursive expansion, this can be expressed as
\begin{align}
U_m&=-2^{m}J\mathrm{Tr}(D_1g_{c1}(1+D_1K_1))^m\nonumber\\
&=-2^{m}J\mathrm{Tr}(D_1g_{c1}(1+D_1G_1)(1+D_2K_2))^m\nonumber\\
&=\ldots\nonumber\\
&=-2^{m}J \mathrm{Tr}R^m
\end{align}
where
\begin{equation}  
R=D_1g_{c1}(1+D_1G_1)(1+D_2G_2)\ldots(1+D_{r_{max}}G_{r_{max}}).
\label{e:r}
\end{equation}
This is sensible since it includes all the features of the ground-state calculation. It is also exact.

We can imagine coloring plaquettes black and white; like a chess board. Matrices $D_r$ and $G_r$ are color diagonal for even $r$: otherwise off-diagonal. Since $g_{c1}$ is color diagonal, $R$ is off-diagonal and it follows that $U_m=0$ for odd $m$. 
This explicitly excludes any $2J$ excitations involving a finite number of spins.

The specific heat per spin can be derived in terms of the internal energy as
\begin{equation}
c_v=\frac{1}{L^2}\frac{dU}{dT}=\frac{1}{L^2}\left(\frac{4J}{kT^2}\right)\sum_{m=1}^\infty me^{-4Jm/kT}U_{2m}.
\end{equation}
We denote the degeneracy of the $i^{th}$ excited state as $M_i$. Expanding $\ln Z$, we obtain, for example, $U_2=4J(\frac{M_1}{M_0})$, $U_4=8J(\frac{M_2}{M_0} - \frac{1}{2}(\frac{M_1}{M_0})^2)$, and $U_6=12J(\frac{M_3}{M_0} - \frac{M_2}{M_0} \frac{M_1}{M_0} + \frac{1}{3}(\frac{M_1}{M_0})^3)$.

\begin{figure}[h]
\includegraphics*[angle=-90,width=8.5cm]{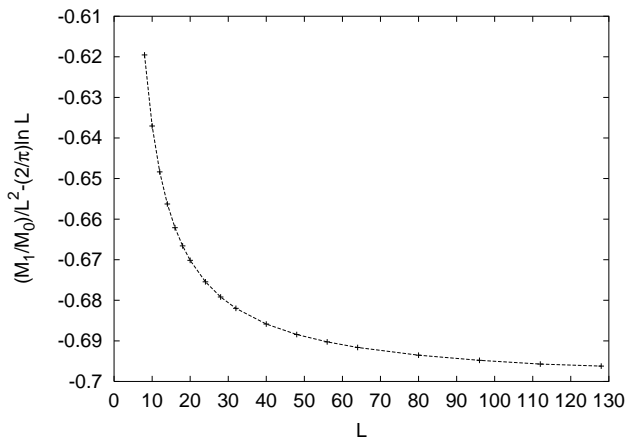}
\caption{\label{f:Fig1} The degeneracy of the first excitations for the
Villain model as a function of lattice size $L$. The curve is a guide for
the eye.}
\end{figure}

As a means of establishing bearings, we first report a study of the first
excitations of the fully frustrated Villain model \cite {Villain,Andre}.
Fig. \ref{f:Fig1} shows clearly that the ratio
$\frac{M_1}{M_0} \sim \frac{2}{\pi} L^2 \ln L$. The correlation length
is known to be $\xi \sim \delta^{-1}$ \cite {Europhys}. Replacing $\ln L$
with $-\ln \delta \sim T^{-1}$, the correct form for the specific heat is
obtained, that is $c_v \sim T^{-3} \exp(-4J/kT)$. The degeneracy per spin of the
first excited state is infinite in the thermodynamic limit, although only
weakly (logarithmically) so.

\begin{figure}[h]
\includegraphics[angle=-90,width=8.5cm]{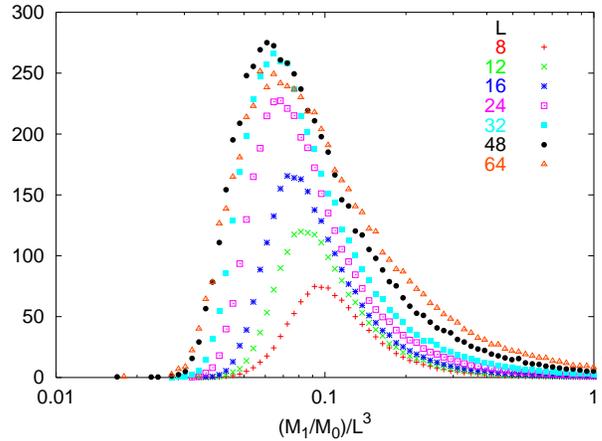}
\caption{\label{f:Fig2}(color online). Distributions, scaled with a factor $L$,
for $\frac{M_1}{M_0}\frac{1}{L^3}$ with various values of lattice size $L$.
Each distribution includes $10^5$ disorder realizations.}
\end{figure}

In Fig. \ref{f:Fig2} we show the distribution of $\frac{M_1}{M_0}$ for the
spin glass. It is clear that the most likely value scales as $L^3$. In this
case we have an extra factor $L$, unlike $\ln L$ for the Villain model. 
In consequence, taking $\xi \sim \exp(2J/kT)$, the specific heat varies
like $c_v \sim T^{-2} \exp(-2J/kT)$ explaining why a $2J$ energy gap arises
from the form of the correlation length obtained from Monte Carlo
calculations \cite {H01,KL04}. Hyperscaling fails here. We also emphasize that the Villain model is not a spin glass and comparisons, such as in Ref. \cite{SK93}, are not meaningful.     

The distributions are scaled with a factor of $L$, probably indicating that
the amount of effort required for an experiment to find the mode of the
distribution scales like $L$. This is consistent with the known difficulties
that arise when trying to extrapolate Monte Carlo data to zero temperature.
Also, the distributions are obviously neither normal or lognormal.
Further, the data is not self averaging. In fact, the relative variance
grows quickly with $L$ which is unusually severe; convergence to a
constant is normally expected \cite {Wiseman}.

We can discuss our choice of correlation length in the light of Ref.
\cite {KLC07}. Two crossover temperatures are defined. Below
$T^{*}(L)$ the ground state behaviour dominates. Above the finite-size
crossover temperature $T_{\xi}(L)$ no size limitations are expected.
It is clearly stated that, in all situations, $T_{\xi} > T^{*}$. This
immediately rules out $\xi \sim \exp(J/kT)$. For our case,
we can in fact make the definite prediction $T_{\xi} = 1.5 T^{*}$.
A power law behaviour for $\xi$ must also be ruled out if we also
expect similar behaviour for the specific heat.

Pfaffians are also computed, for the partition function, in Ref.
\cite {LGM04} at definite finite temperatures. For comparison, we have
computed the mean of $\ln(\frac{M_1}{M_0})$ and find fairly close agreement.
Nevertheless, we do not believe that this is physically meaningful. The
distributions of $\frac{M_1}{M_0}$ 
are not lognormal and the most likely value scales as $L^3$, not $L^4$. We emphasize here that we are not computing the entire Pfaffians.

\begin{figure}[ht]
\includegraphics[angle=-90,width=8.5cm]{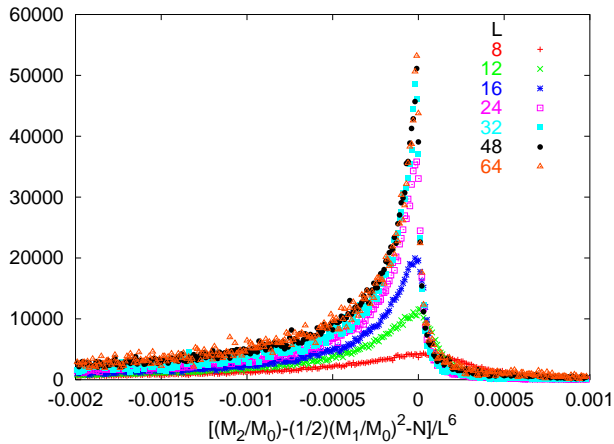}
\caption{\label{f:Fig3} (color online). Distributions, scaled with a factor $L$, for the second term in the specific
heat. Each includes $10^5$ disorder realizations, except for $L=64$ which has $35,000$.}
\end{figure}
\begin{figure}[ht]
\includegraphics[angle=-90,width=8.5cm]{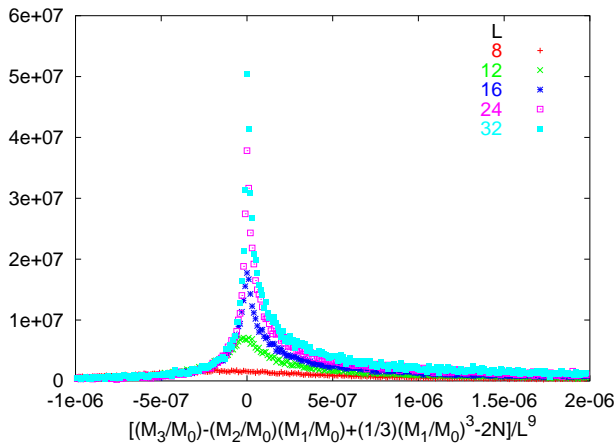}
\caption{\label{f:Fig4} (color online). Distributions, scaled with a factor $L$, for the third term in the specific
heat. Each includes $10^5$ disorder realizations.}
\end{figure}

We have also checked distributions for second excitations. The most likely
value of $\frac{M_2}{M_0}-N$ scales as $L^6$. We have subtracted off here the (infinite) total number of spins $N$. Single spin flips can occur anywhere and it is sensible to measure the
internal energy relative to the unfrustrated system. However, the appropriate contribution to the internal energy is
$\frac{M_2}{M_0} - \frac{1}{2} (\frac{M_1}{M_0})^2 - N$. As shown in
Fig. \ref{f:Fig3}, the most likely value is close to zero. We note that the mode of $\frac{1}{2}(\frac{M_1}{M_0}\frac{1}{L^3})^2$ is roughly $0.002$; indicating a close cancellation. Note also that the shapes of the distributions show clearly the dominant effect of the first excitations. The distributions for third excitations, that is of
$\frac{M_3}{M_0} - \frac{M_2}{M_0} \frac{M_1}{M_0} + \frac{1}{3} (\frac{M_1}{M_0})^3 -2N$,
also show a most likely value close to zero as shown in Fig. \ref{f:Fig4}. Here, the mode of $\frac{M_3}{M_0}-2N$ scales as $L^9$ and the unfrustrated system has $M_3=2N$. It is unlikely that higher
excitations will interfer with our arguments based on first excitations.

In conclusion, we have reported exact results for the excitations of the
bimodal two-dimensional Ising spin glass by expanding in arbitrary
temperature from the ground state. All other treatments, except Ref. \cite{SK93}, have required extrapolation from definite finite temperatures. We have argued that
an energy gap of $2J$ is consistent with $\xi \sim \exp(2J/kT)$ as found
from Monte Carlo simulations. The manner in which our model is arranged
excludes the possibility of any $2J$ excitation for a finite system.
Nevertheless, it should also be interesting to study systems with obvious $2J$ excitations, for example the hexagonal lattice or square lattice with open boundaries or diluted bonds, and investigate the distributions.

W. A. thanks the Commission on Higher Education Staff Development Project, Thailand for a scholarship. J. P. acknowledges fruitful conversations with J. A. Blackman. Some of the computations were performed on the Tera Cluster at the Thai National Grid Center.


\begin{references}
\bibitem{bond_basis}
It is convenient to use the bond basis $\mid \pm\rangle=\frac{1}{\sqrt{2}}(\mid \alpha\rangle\pm\zeta_{ij}\mid\beta\rangle)$ where $\mid \alpha\rangle$ and $\mid \beta\rangle$ are as shown in Fig. 4. of \cite{BP91}.
\end{references}
\end{document}